\documentclass[12pt]{iopart}
\usepackage{graphics}
\usepackage{epsfig}
\begin{document}
% Title Page
\title{Fragment Formation in Biased Random Walks}
\author{Kabir Ramola}
\address{Department of Theoretical Physics\\
Tata Institute of Fundamental Research, Mumbai}
\ead{kabir@tifr.res.in}

\begin{abstract}
We analyse a biased random walk on a 1D lattice with unequal step lengths. Such a walk was recently shown to undergo a phase transition from a state containing a single connected cluster of visited sites to one with several clusters of visited sites (fragments) separated by unvisited sites at a critical probability $p_{c}$ [\textit{PRL \textbf{99}, 180602 (2007)}]. The behaviour of $\rho(l)$, the probability of formation of fragments of length $l$ is analysed. An exact expression for the generating function of $\rho(l)$ at the critical point is derived. We prove that the asymptotic behaviour is of the form $\rho(l) \simeq 3/[l(\log\ l)^2]$.
\end{abstract}
\maketitle

\section{Introduction}
Anteneodo and Morgado recently discussed an interesting one-dimensional random walk model that exhibits a phase transition \cite{Anteneodo}. In this model, at each time step the random walker moves two lattice spacings to the right or one spacing to the left with probabilities $p$ and $q$ respectively. The sites visited by such a walk are, depending on $p$, either part of a single connected cluster of visited sites or many clusters of visited sites (fragments) separated by single unvisited sites. A transition from one class to the other takes place at the critical probability $p_c=1/3$. At large times, for $p<p_c$ the cluster of visited sites contains no gaps, however, for $p>p_c$ this cluster contains a finite density of unvisited sites. Using Monte-Carlo simulation data, a power law dependence with an exponent $\simeq 1.15$ was estimated for $n(l)$, the fraction of fragments with length $l$. Since the model is a simple one dimensional random walk that is not known to contain any such anamolous exponents, such a behaviour seems unlikely. We prove below that the correct asymptotic behaviour of $n(l)$ at the critical point is $n(l)\sim [l({\textmd{log}}\ l)^2]^{-1}$ which is infact hard to distinguish numerically from a power law dependence with exponent $\simeq 1.15$. 

At large times, it is easy to see (using the central limit theorem) that the probability distribution of the position of the walker is a Gaussian with mean  $\mu(t)=(2p - q)t$ and variance $\sigma^{2}(t)=9pqt$. When the walker is biased towards the left ($p < 1/3$), the probability of the walker being on the left of the origin tends to $1$ as $erf(\sqrt{t})$. Hence the visited sites are part of a single connected cluster. When the walker is biased towards the right, in general the clusters of visited sites are separated by single unvisited sites. The fraction of visited sites $f_{v}$ is defined as $S_{n}/L_{n}$ where $S_{n}$ is the average number of distinct sites visited in an $n$-step walk and $L_{n}$ is the average length of an $n$-step walk. The fraction of unvisited sites $f_{u}$, is given by $1-f_{v}$. For this case, $f_{u}$ can be calculated exactly using methods outlined in \cite{Montroll},\cite{Anteneodo}. In the large $n$ limit, as $\Delta\rightarrow 0^{+}$ , $f_{u} = \Delta + O(\Delta^{2})$, where $\Delta = (p -p_{c})$. The total number of fragments in a walk is equal to the number of unvisited sites and hence $\overline{N}$, the average number of fragments is equal to $L_{n}\Delta$.

The single unvisited sites at the edge of each fragment ensure that the probability of formation of a fragment is independent of the previous history of the walker. The walk can therefore be considered as a discrete process of independent increments where at each step a new fragment of length $l$ is added to the positive edge of the walk with probability $\rho(l)$. These probabilities are multiplicative, i.e.- the probability of formation of a fragment of length $l_{1}$ followed by one of length $l_{2}$ is $\rho(l_{1})\rho(l_{2})$. The number of fragments of length $l$ in a given walk is $N \rho(l)$ where $N$ is the total number of fragments. Therefore $n(l)= \rho(l)/\sum_{l=1}^{\infty} \rho(l)$ and hence $n(l) = \rho(l)$ since $\sum_{l=1}^{\infty} \rho(l)= 1$.

\section{Determination of fragment formation probability}
For a new fragment of length $l$ to be formed the walker must (i) be at the positive edge of the walk ( at site $-1$ for convenience), (ii) hop over site $0$ and reach site $l$ by visiting each site between $1$ and $l$, without visiting sites $0$ and $l+1$. Thus at the end of each step the walker is once again at the positive edge of the walk. $\rho(l)$ is therefore the probability of event (ii) occuring. Hence $\rho(l)$ is the sum of probabilities of all paths that are consistent with (ii). 

We calculate $q(l)$, the probability of condition (ii) with the constraint of `visiting every site between $1$ and $l$' relaxed. Hence $q(l)$ is the probability of starting at site $-1$ and reaching site $l$ without visiting $0$ and $l+1$. $q(l)$ thus includes the probability of formation of smaller fragments within the segment $[1,l]$. For example $q(3) = \rho(3) + \rho(1)\rho(1)$. In general
\begin{equation}
q(l) = \rho(l) + \sum_{x_{1}+x_{2}+1=l}^{}\rho(x_{1})\rho(x_{2}) + \sum_{x_{1}+x_{2}+x_{3}+2=l}^{}\rho(x_{1})\rho(x_{2})\rho(x_{3}) + ...
\label{generalq}
\end{equation}
This translates to the following equation involving generating functions
\begin{equation}
\overline{q}(x)= \frac{\overline{\rho}(x)}{1-x\overline{\rho}(x)}, \textmd{~~~i.e.-~~~~} \overline{\rho}(x)= \frac{\overline{q}(x)}{1+x\overline{q}(x)} \\
\label{generating}
\end{equation}
where 
$\overline{\rho}(x)=\sum_{l=1}^{\infty}\rho(l)x^{l}$ and  $\overline{q}(x)=\sum_{l=1}^{\infty} q(l)x^{l}$.\\

\subsection{Exact expression for $q(l)$}
We calculate $q(l)$ by summing over probabilities of all paths that are consistent with the above definition of $q(l)$. For example, $q(1)$ simply involves the walker reaching site $1$ from $-1$ by hopping over site $0$. Thus $q(1)=p$. It is also easy to see that $q(2)$ is $0$ since unvisited sites in this walk cannot be separated by two lattice sites. In the calculation of $q(3)$ we sum over paths in which the walker starting from $-1$, jumps over site $0$ to reach $1$ and then over $2$ to reach site $3$ with probability $p^2$. Once at site $3$, the walker can move two leftward steps followed by a rightward step with probability $pq^{2}$, any number of times, to return to site $3$. Thus $q(3) = p^{2}( 1 + pq^{2} + (pq^{2})^{2} + ...) = p^{2}/(1-pq^{2})$. Generalizing this procedure of summing over all relevant walks, we derive below a closed form expression of $q(l)$ in two equivalent ways.

The walk can be considered as a Markov process where at each time step the walker moves either two lattice spacings to the right or one to the left with probabilities $p$ and $q$ respectively, independent of the previous step. The probability $q(l)$ can be calculated from  $P_{l}$, the matrix of transition probabilities \cite{Feller} with no transitions from the sites $0$, $l+1$ and $l+2$. For example 
\begin{equation}
P_{3}= \left[\begin{array}{cccccc}
0 ~~q ~~0 ~~0 ~~0 ~~0 \\ 
0 ~~0 ~~q ~~0 ~~0 ~~0 \\
0 ~~0 ~~0 ~~q ~~0 ~~0 \\
0 ~~p ~~0 ~~0 ~~0 ~~0 \\
0 ~~0 ~~p ~~0 ~~0 ~~0 \\
0 ~~0 ~~0 ~~p ~~0 ~~0 \\
               \end{array}\right]
\end{equation}
Where the rows correspond to sites [0 1 2 3 4 5]. Consider the column vector $\vert p(t)\rangle$ where $\langle s \vert p(t) \rangle$, the element in row $s$, is the probability of the walker starting at site $1$, being at site $s$ at time step $t$. Now $\vert p(t)\rangle$ evolves as $\vert p(t)\rangle = P_{l}^{t}\vert p(0) \rangle$ which ensures that $\vert p(t)\rangle$ only contains probabilities of paths that never visit sites $0$ and $l+1$. Since the walker starts at site $1$, $\langle s \vert p(0) \rangle = \delta_{1,s}$. We calculate $\vert P \rangle$ where $\langle s \vert P \rangle$ is the probability of the walker, starting from site $1$ being at site $s$, at any time step. $\vert P \rangle$ is thus the sum of $\vert p(t)\rangle$ at every time step. Hence
\begin{equation}
\vert P \rangle =(I + P_{l} + P_{l}^{2} + ...) \vert p(0)\rangle
\end{equation}
All walks that start at $1$ and end at $l$ (at any time step) contribute to $q(l)$, therefore 
\begin{equation}
q(l)= p \langle l \vert (I-P_{l})^{-1}\vert 1 \rangle
\label{inversion}
\end{equation}
where $\langle s \vert 1 \rangle = \delta_{1,s}$ and $\langle s \vert l \rangle = \delta_{l,s}$. Thus $q(l)$ is the ($l,1$) matrix element of $p(I - P_{l})^{-1}$. The factor $p$ is present because the walker must hop over site $0$ to reach $1$ as in condition (ii).  
We thus obtain $q(1)= p$, $q(2)= 0$, $q(3)= \frac{p^{2}}{1-pq^{2}}$, $q(4)= \frac{p^{3}q}{1-2pq^{2}}$ and so on. Subtituting these into Eq.(\ref{generating}) we obtain  $\rho(1)= p$, $\rho(2)= 0$, $\rho(3)= \frac{p^{3}q^{2}}{1-pq^{2}}$, $\rho(4)= \frac{p^{4}q^{2}+3p^{5}q^{4}}{(1-pq^{2})(1-3pq^{2})}$ and so on.

Alternatively, consider $G(s)$, the probability of a walker starting at site -1, being at site $s$ while never visiting $0$ and $l+1$. Now $G(s)$ satisfies the difference equation
\begin{equation}
G(s)=pG(s-2) + qG(s+1)
\label{gs}
\end{equation}
with the boundary conditions $G(0)=0$ and $G(l+1)=0$. A general solution of $G(s)$ is $Az_{1}^{s} + Bz_{2}^{s} + Cz_{3}^{s}$ which on substitution into Eq.(\ref{gs}) yields the cubic equation $qz^{3}-z^{2}+p=0$. The solutions to this equation are $z=1, z_{\pm}$ where  $z_{\pm}=(p\pm\sqrt{p^{2}+4pq})/2q$. By satisfying boundary conditions we obtain 
\begin{equation}
G(s)= A[1 -\frac{(1-z_{-}^{l+1})}{z_{+}^{l+1}-z_{-}^{l+1}}z_{+}^{s}-\frac{(z_{+}^{l+1}-1)}{z_{+}^{l+1}-z_{-}^{l+1}}z_{-}^{s}]
\label{gs-solution}
\end{equation}
where A is a constant which is determined from the normalization condition $G(-1)=1$. Alternatively from the definition of $q(l)$ as the connecting probability between $-1$ and $l$ we have 
\begin{equation}
q(l)=G(l)/G(-1)
\label{qsolution}
\end{equation}

We thus have a closed form expression for $q(l)$. We verify that the two expressions for $q(l)$ are equal as the matrix elements of the column corresponding to site 1 in $p(I-P_{l})^{-1}$ satisfy the same difference equation and, with an appropriate change of variables, the same boundary conditions as G(s).

\begin{figure}[!ht]
\begin{center}
\epsfig{file=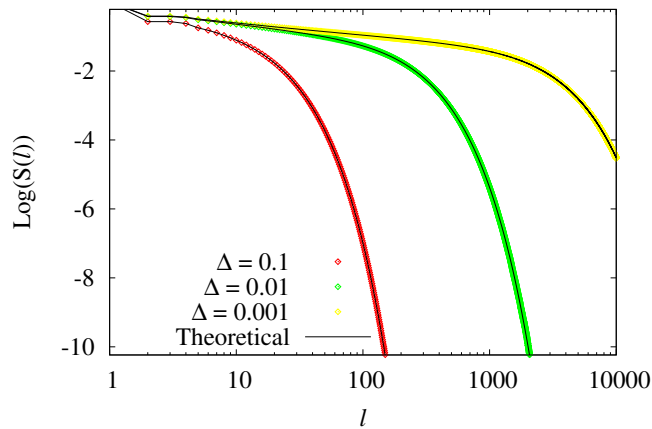,height=6cm}
\caption{Monte-Carlo plots of $\log(S(l))$  where $S(l)= \sum_{\geq l} n(l)$ for different values of $\Delta = p-p_{c}$ , along with the theoretical curve calculated using Eq.(\ref{qsolution}) and Eq.(\ref{generating}). }
\end{center}
\end{figure}

\section{Asymptotic behaviour}
Away from the critical point $(p>p_{c})$, for large $l$ 
\begin{equation}
q(l) \simeq \frac{1-\frac{1}{z_+}}{1-\frac{1}{z_-}}-\frac{z_{-}(z_{+}-z_{-})(z_{+}-1)}{(1-z_{-})^{2} z_{+}^{3}}\textmd{Exp}[-l/\xi]\\
\label{noncriticalq}
\end{equation}
where $\xi =\frac{1}{\log [z_{+}]} \simeq \frac{1}{3(p-p_{c})}$. $\xi$ thus defines a correlation length for $q(l)$ and hence for $\rho(l)$.\\

At the critical point $q(l)$ simplifies to
\begin{equation}
q(l)=\frac{2[1-(-\frac{1}{2})^l]-3l(-\frac{1}{2})^l}{8 + (-\frac{1}{2})^l + 6l}\simeq\frac{1}{3l} \textmd{~~~for large $l$}
\label{criticalq}
\end{equation}
Therefore as $x\rightarrow1$ the generating function $\overline{q}(x)$ diverges as $-\frac{1}{3}\log(1-x)$. Hence from Eq.(\ref{generating})
\begin{equation}
\overline{\rho}(x)= \frac{-\frac{1}{3}\log(1-x)}{1-\frac{x}{3}\log(1-x)}\textmd{~~~for ($x\rightarrow1 , p=p_{c}$)}
\label{criticalrho}
\end{equation}

Having found the generating function, we can extract the behaviour of $\rho(l)$ for large $l$. From Eq.(\ref{criticalrho}), $\overline{\rho}(x)\simeq1+3/\log \epsilon$ where $\epsilon=(1-x)$.
We can approximate the summation in the generating function by an integral $\overline{\rho}(x)\simeq\int_{1}^{\infty}\textmd{Exp}[-\epsilon\ l]\rho(l)dl$. The singular part of this integral can be evaluated by equating $\overline{\rho}(x)\simeq\int_{1}^{1/\epsilon}\rho(l)dl$ from which we see that $\int_{1/\epsilon}^{\infty}\rho(l)dl\simeq -3/\log \epsilon$. Alternatively, from the fact that the integral $\textmd{Lim}_{\epsilon\rightarrow0}\int_{2}^{\infty}\textmd{Exp}[-\epsilon\ l]/(\log \ l)^2dl$ diverges as $[\epsilon(\log \epsilon)^2]^{-1}(1+O[\frac{1}{\log \epsilon}])$ we obtain
\begin{equation}
\rho(l)\simeq\frac{3}{l(\log l)^{2}}\textmd{~~~for large $l$}
\label{asymptoticrho}
\end{equation}
We thus have an expression for the asymptotic behaviour of $\rho(l)$. 

Cases in which the rightward steps are of length $k > 2$ can be treated in a similar way. In this case the independent elements are clusters of visited and unvisited sites separated by $k-1$ zeroes.

\begin{figure}[t]
\begin{center}
\epsfig{file=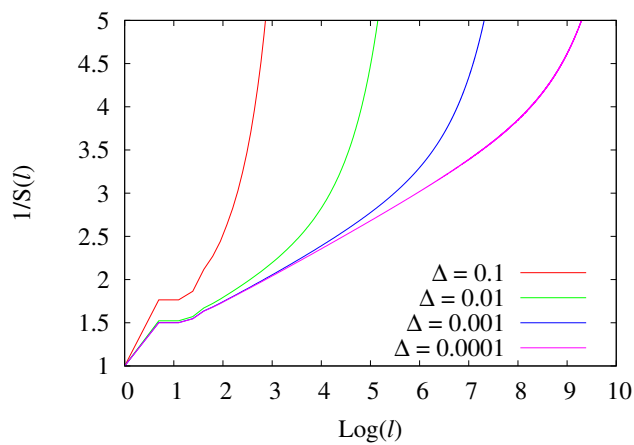,height=6cm}
\caption{Theoretical plots of $1/S(l)$ v/s Log$(l)$ for different values of $\Delta$ showing  a limiting slope of $\simeq$ 0.33 in accordance with Eq.(\ref{asymptoticrho})}
\end{center}
\end{figure}

\section{Acknowledgement}
I thank Prof.~Deepak Dhar for providing the central ideas that led to these results and R. Loganayagam for helpful discussions on the series expansions involved.\\

\end{document}